\documentclass[sigconf]{acmart}

\usepackage{afterpage}

\copyrightyear{2022} 
\acmYear{2022} 
\setcopyright{acmlicensed}\acmConference[ICSE '22 Companion]{44th International Conference on Software Engineering Companion}{May 21--29, 2022}{Pittsburgh, PA, USA}
\acmBooktitle{44th International Conference on Software Engineering Companion (ICSE '22 Companion), May 21--29, 2022, Pittsburgh, PA, USA}
\acmPrice{15.00}
\acmDOI{10.1145/3510454.3516837}
\acmISBN{978-1-4503-9223-5/22/05}

\begin{document}

\title{Proactive Libraries: Enforcing Correct Behaviors in Android Apps}

\author{Oliviero Riganelli}
\email{oliviero.riganelli@unimib.it}
\affiliation{
  \institution{University of Milano - Bicocca}
  \city{Milan}
  \country{Italy}
}
\author{Ionut Daniel Fagadau}
\email{i.fagadau@campus.unimib.it}
\affiliation{
  \institution{University of Milano - Bicocca}
  \city{Milan}
  \country{Italy}
}
\author{Daniela Micucci}
\email{daniela.micucci@unimib.it}
\affiliation{
  \institution{University of Milano - Bicocca}
  \city{Milan}
  \country{Italy}
}
\author{Leonardo Mariani}
\email{leonardo.mariani@unimib.it}
\affiliation{
  \institution{University of Milano - Bicocca}
  \city{Milan}
  \country{Italy}
}

%
%
%
%
%
%
%

\renewcommand{\shortauthors}{O. Riganelli, I. D. Fagadau, D. Micucci, and L. Mariani}

\begin{abstract}

The Android framework provides a rich set of APIs that can be exploited by developers to build their apps. However, the rapid evolution of these APIs jointly with the specific characteristics of the lifecycle of the Android components challenge developers, who may release apps that use APIs incorrectly.

In this demo, we present \emph{Proactive Libraries}, a tool that can be used to decorate regular libraries with the capability of proactively detecting and healing API misuses at runtime. \emph{Proactive Libraries} blend libraries with multiple proactive modules that collect data, check the compliance of API usages with correctness policies, and heal executions as soon as the possible violation of a policy is detected.
The results of our evaluation with 27 possible API misuses show the effectiveness of \emph{Proactive Libraries} in correcting API misuses with negligible runtime overhead.
\\
\textbf{Video}: \href{https://youtu.be/rkfZ38mPgV0}{https://youtu.be/rkfZ38mPgV0}\\
\textbf{Repo}: \href{https://gitlab.com/learnERC/proactivelibrary}{https://gitlab.com/learnERC/proactivelibrary}
\end{abstract}




\maketitle
\afterpage{
 \begin{figure*}[ht!]
\begin{center}
  \includegraphics[width=\textwidth]{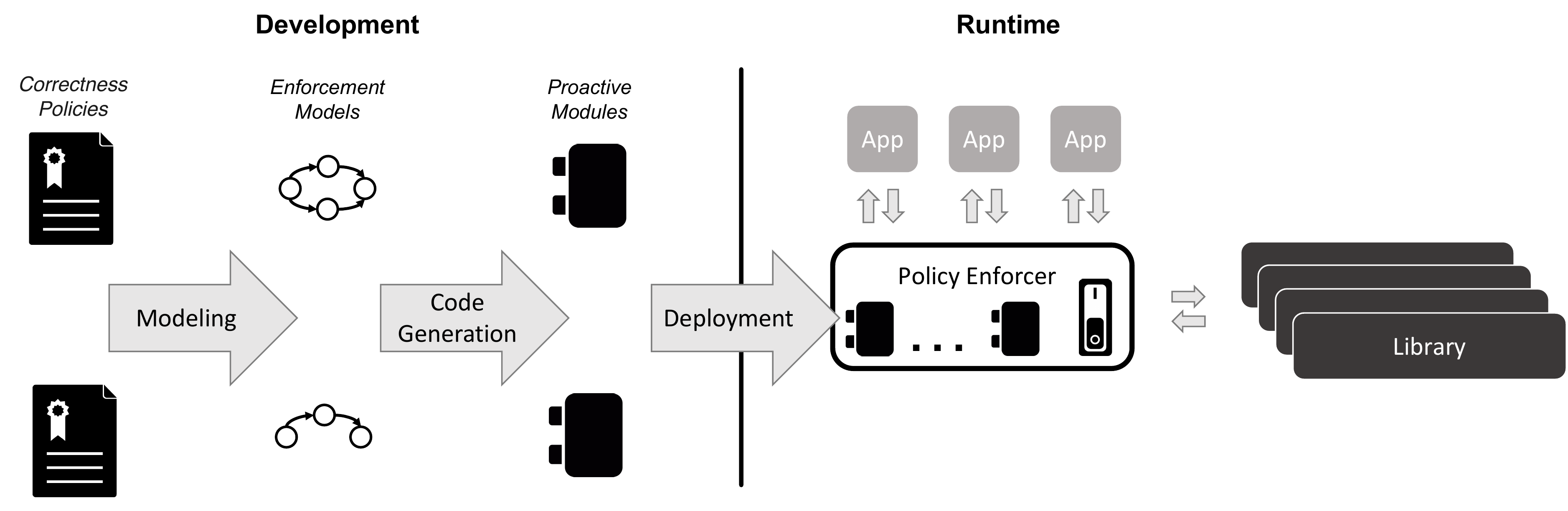}
\caption{Proactive Libraries.}
\label{fig:proactivelibrary}
\end{center}
\end{figure*}
}
\section{Introduction}\label{sec:introduction}
Users of mobile devices can easily access to a huge number of apps through well-established digital marketplaces\footnote{\url{https://play.google.com/store}},\footnote{\url{https://www.apple.com/app-store}}. 
Marketplaces are populated with apps ranging from professional ones to those developed by hobbyists. The openness of the marketplaces facilitates software distribution but also raises relevant reliability issues. In fact, many of these apps are often not thoroughly validated and may cause troubles once installed and executed in a device.

Users regularly report issues with the apps downloaded from the marketplaces~\cite{Azim_Towards_2014,Banerjee:EnergyBugs:FSE:2014,Wei:AndroidFragmentation:ASE:2016,Shan:ResumerRestartErrors:OOPSLA:2016,Wu:Callback:TSE:2016}. A relevant portion of the problems experienced by the users  is related to the way apps use APIs and specifically how they are used within the lifecycle of an event-driven Android object \cite{Wang:APIObstacles:2013,Su:AppCrashes:TSE:2020}. Indeed, API misuses are pervasive and are the cause of many software errors, failures, and vulnerabilities~\cite{Amani:MUBench:MSR:2016,Monperrus:MissingMethopdCalls:TOSEM:2013,Amann:APIMisuseDetectors:TSE:2019,Nadi:CryptographyAPI:ICSE:2016}.
For instance, many apps fail to properly acquire and release resources (e.g., the camera and  the microphone) within the lifecycle of an activity\footnote{Android apps are composed of multiple components called activities: \url{https://developer.android.com/guide/components/activities}}, causing efficiency and energy problems~\cite{Azim_Towards_2014,Banerjee:EnergyBugs:FSE:2014,Wu:Callback:TSE:2016}.  In some cases, inaccurate interactions with the APIs managing the resources may even cause problems across apps. For instance, an app that does not release the camera every time its execution is suspended may prevent the other apps from acquiring the same camera~\cite{Azim_Towards_2014,Wu:Callback:TSE:2016}.

The incorrect interaction between an app and a resource can be often recognized by looking at both the usage of the API that controls the access to the resource and the lifecycle of the components of the app that interacts with the library (e.g., the lifecycle of an Activity). For example, if the user moves to background an app that is using the microphone to record audio, the app must immediately release the microphone, otherwise the other apps might be unable to interact with the microphone because it is held by the app running in the background. Moreover, 
the device may unnecessarily keep the microphone active consuming extra battery. At the level of the interaction between the app and the library this means that once an app has invoked the method \texttt{startRecording()} of the Android library class \texttt{AudioRecord}, it must release the microphone by invoking the method \texttt{release()} every time a call to the \texttt{onStop()} callback method is generated by the Android framework\footnote{\texttt{onStop()} is a callback method that apps implement to define the operations that must be performed before an activity becomes invisible to the user}. Unfortunately, it is not always the case that apps are implemented releasing and acquiring resources coherently with the lifecycle of the activities~\cite{Liu:ResourceLeaks:ISSRE:2016,Azim_Towards_2014,Wu:Callback:TSE:2016}.

To prevent users from experiencing annoying problems caused by apps that inaccurately interact with APIs, we introduced a novel concept of library, called \textit{Proactive Library}~\cite{Riganelli:TAAS:PL:2019,Riganelli:ProactiveLibraries:SEAMS:2017}, that forces the apps to behave as the APIs require. 
For example, Proactive Library forces the app to release the camera every time its execution is suspended, that is, every time the \texttt{onPause()} callback of the activity that acquires the camera is invoked.

A proactive library is composed of two parts: the reactive library, which is a regular implementation of an API, and a set of runtime \emph{enforcers}, called \emph{proactive modules}, which decorate the library with the ability to enforce some \emph{correctness policies} at runtime. 

The proactive modules operate jointly with their library reacting to the invocation of certain API and callback methods, checking that the apps are using the API appropriately, and automatically fixing the execution and the status of the system, if necessary. For example, when the \texttt{onStop()} callback is generated, the proactive module of the library for audio recording is triggered for checking if the microphone has been released. If not, the proactive module may force the release of the microphone, and automatically reassign it again to the activity once the activity becomes visible to the user again. To make the solution broadly applicable, proactive modules do not require access to the source code of the app, but work in a black-box fashion. In fact, any developer may design a proactive module for a library and distribute it. 

Proactive modules can be installed and speculatively activated or deactivated by smartphone users via a dedicated app. For instance, proactive modules might be activated by users after some apps presented some misbehaviours. 

To support development activities, we defined a model-based approach for the definition of the proactive modules and the automatic generation of their implementation for the Android environment. This paper describes the supporting tool we developed to design, implement, and activate proactive libraries. 
We used the tool with 27 possible API misuses in real Android apps showing that Proactive Libraries can effectively correct library misuses with negligible runtime overhead.
 
 This paper is organized as follows. Section \ref{sec:approach} describes tool support for \textit{Proactive Libraries}. Section~\ref{sec:evaluation} presents our empirical evaluation. Section~\ref{sec:related} discusses related work. Section~\ref{sec:conclusion}  provides final remarks.

\section{Proactive Libraries}\label{sec:approach}
Figure~\ref{fig:proactivelibrary} visually illustrates how Proactive Libraries work, distinguishing the development and the runtime phases. Our tool implementation supports both phases.

At development time, developers of proactive modules first identify the correctness policies that they want to enforce. A \emph{correctness policy} is a natural language statement obtained form the API documentation that constrains the usage of the API, considering the status of the app as captured by its lifecycle, if necessary. Any app that violates a correctness policy is faulty. An example of a correctness policy related to resource usage for the \texttt{Camera} Android library is:


\begingroup
\addtolength\leftmargini{-0.1in}
\begin{quote}
``\emph{An activity that is paused while having the control of the camera must first release the camera}.''\end{quote}
\endgroup

The policy refers to both the status of the activity (an activity that is paused) and the usage of the API, and it is valid for every app that uses the camera. 

Correctness policies are then formalized and encoded as enforcement models. An \emph{enforcement model}  defines how to react to any violation of the correctness policies. We use edit-automata to specify enforcement models since they are simple finite-state based formalizations that naturally support the specification of the behavior of the proactive module in terms of the events that must be intercepted and the events that must be inserted and suppressed as a reaction to each intercepted event. 
The objective of the enforcement model is to \emph{enforce} a correctness policy, when the running apps do not satisfy it. The enforcement model is defined uniquely using the knowledge of the library API and the Android lifecycle events (i.e., the callback methods)~\cite{Android:Lifecycle:website}, which are the same for any app. Thus its definition does not require any knowledge related to the app that uses the library. 

An enforcement model fully describes the behavior of the corresponding proactive module, which can be thus generated from the model. Proactive modules can be deployed in any environment where the corresponding library is used. Since proactive modules are activated by the invocation of certain methods, their execution in the user environment is controlled by a \emph{policy enforcer} that intercepts the events and dispatches them to the deployed proactive modules. The policy enforcer also controls the activation and deactivation of the proactive modules. 

\subsection{Enforcement Models Definition}
An enforcement model is a formal representation of the actions that must be undertaken to automatically enforce a correctness policy, that is, the operations that must be executed to turn an execution that violates a policy into an execution that satisfies it.  
Our tool lets proactive module developers visually specify enforcement models as edit automata~\cite{Ligatti:Edit:JIS:2005} directly from within the Eclipse IDE. Our implementation focuses on the Android ecosystem and on a specific class of correctness policies, the \emph{resource usage policies}, which state how an API that controls the access to a resource must be used by apps. An interesting aspect about these policies is that the usage of a resource is strongly coupled with the status of the apps and their components, that is, depending on the state of an Android component as captured by its lifecycle there are operations that must or must not be executed. These two elements, the API that handles the target resource and the Android component that interacts with the API, must be explicitly identified through their class names every time a new enforcer is defined with our tool. This information enables the possibility to later generate code that traces the state of the Android component by intercepting its  callbacks, 
and the state of the API managing the resource by intercepting the calls to the API methods.

Callback methods are method calls produced by the Android framework when an app changes its status. For instance, the callback method \texttt{onStop()} is automatically invoked when the running activity of the app is no longer visible, while \texttt{onDestroy()} is invoked when the activity is destroyed. 
The API methods are the methods implemented by the library associated with the proactive module. For instance, in the case of the camera, the methods \texttt{open()} and \texttt{release()} are API methods implemented by the \texttt{Camera}. 

The edit automata also include the actions that must be proactively and automatically suppressed or inserted to enforce the satisfaction of a possibly violated policy. 
Again, the calls that can be added and/or suppressed are callback methods and API calls.

%

\begin{figure}
\begin{center}
  \includegraphics[width=0.45\textwidth]{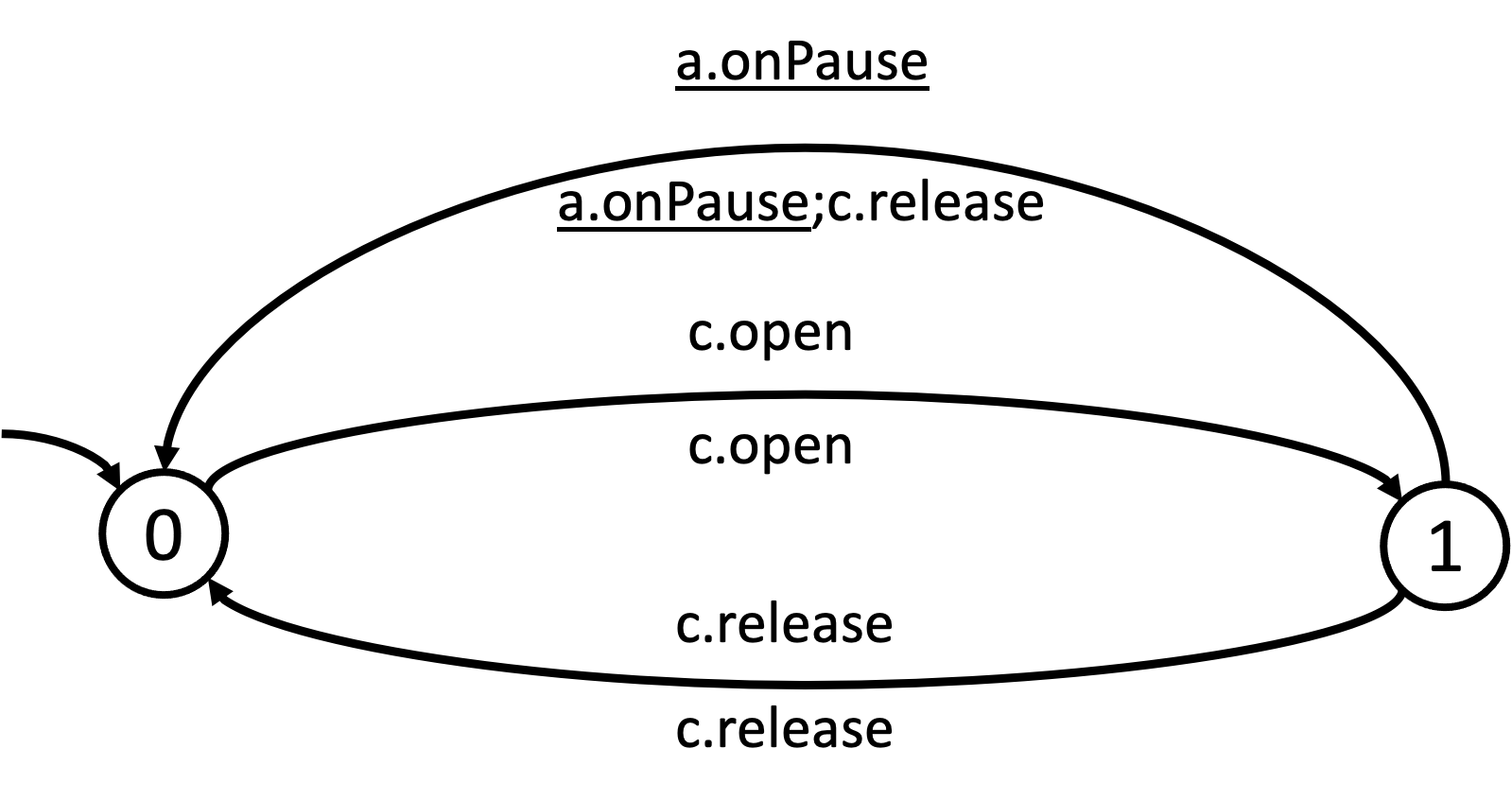}
\caption{Enforcement model for the Resource Usage Policy 1 of the \texttt{Camera}.}
\label{fig:simplifiedmodel}
\end{center}
\end{figure}

Figure~\ref{fig:simplifiedmodel} shows the enforcement model (i.e., an edit automaton) for the example policy about the \texttt{Camera} API reported in Section~\ref{sec:approach}. The symbol above a transition indicates the input symbol recognized by the transition, while the sequence below a transition indicates the output sequence emitted by the automaton when the input sequence is recognized, that is, the model specifies how the execution must be modified step-by-step. 

Compared to general edit automata, our tool extends the standard notation to support two aspects specific to our application context: prefixes and typed events. 

We use \emph{prefixes} to disambiguate the element that performs the operations represented in the model. In this example, the prefix \texttt{a} represents the Activity class, and the prefix \texttt{c} represents the camera API. In the actual automata editor the full name of classes is used as prefix of methods. 

The \emph{events} in the model could be of two types: events that identify the beginning of an operation and events that identify the end of an operation. The type of an event can be distinguished based on its prefix: prefix \texttt{after\#} corresponds to the end of an operation, while prefix \texttt{before\#} corresponds to the beginning of an operation. For example, the outgoing transition from state $0$ in Figure~\ref{fig:simplifiedmodel} represents the start of the execution of the API method \texttt{open}, while the transitions from state $1$ to state $0$ represent the completion of the execution of the \texttt{open} and \texttt{onPause} methods belonging to the camera and activity classes respectively. This distinction is important to precisely represent the events that must activate the enforcers. For instance,  the enforcer must capture the end of the execution of the callback method \texttt{onPause} originated from state $1$ to detect that the camera has not been released while executing \texttt{onPause}.

The example model in Figure~\ref{fig:simplifiedmodel} enforces the release of the \texttt{Camera} when the activity is paused while it still has the camera. In the initial state (state $0$), the enforcer checks whether a camera is opened.  If a camera is opened, the enforcer checks that it is released before the activity is paused (transition from state $1$ to state $0$ with \texttt{release} as input symbol). 
If an app uses the camera consistently with the policy, the execution is unaltered. However, an app might be paused without releasing the camera (transition from state $1$ to state $0$ with \texttt{onPause} as input symbol). In this case the enforcer changes the execution forcing the release of the camera. 

\subsection{Tool Support}
Since implementing proactive modules manually is an error prone and time consuming process, our tool implementation supports a Model-Driven Software Development (MDSD) process and the corresponding tool chain to obtain proactive modules automatically from the enforcement models.
Our MDSD tool chain implements the four core functionalities of a MDSD environment:
\begin{itemize}
\item \emph{Modeling}, which provides the capability to model enforcement models that encode resource usage policies;
\item \emph{Validation}, which provides the capability to validate the syntactic correctness of an enforcement model;
\item \emph{Code generation}, which provides the capability to automatically generate the source code that implements the enforcement logic described by the enforcement model; 
\item \emph{Compilation} and \emph{deployment}, which provide the capability to compile and deploy the generated proactive module on a target platform.
\end{itemize}

We assigned these responsibilities to two main components: the \emph{EMEditor}, which implements \emph{modelling} and \emph{validation} capabilities, and the \emph{PMGenerator}, which implements \emph{code generation}, \emph{compilation}, and \emph{deployment} capabilities.

A developer of proactive modules uses the \emph{EMEditor} to visually design a model that enforces the considered policy. The output of the EMEditor is a validated enforcement model. The developer of proactive modules then provides the generated enforcement model in input to the \emph{PMGenerator} 
to generate and deploy the corresponding proactive module. 
Code generation is based on two sets of rules that distinguish the platform independent and the platform dependent part of the process.

We implemented the EMEditor in Eclipse~\cite{Eclipse} using the Eclipse Modeling Framework (EMF)~\cite{EMF} and the Graphical Modeling Framework (GMF)~\cite{GMF} plugins. EMF provides support to create modeling and code generation tools starting from a data model. GMF provides support to implement graphical editors in Eclipse. We thus created the EMEditor by defining models in EMF and obtaining the corresponding graphical editors with GMF. To automatically generate the proactive module from the enforcement model defined with the EMEditor tool, we used  Eclipse~\cite{Eclipse} augmented with the Acceleo~\cite{Acceleo} plugin, which is an open-source code generator for model-driven development. 

The generated proactive modules are Xposed modules~\cite{Xposed_2016}.\linebreak Xposed allows to cost-efficiently intercept method invocations and change the behavior of an Android app using runtime hooking and code injection mechanisms. Xposed transparently intercepts the method calls to the encapsulated resource and to the Android component with a lifecycle and propagates them to the policy enforcer, which reacts based on the behavior defined in the enforcement model. For instance, it might introduce additional method calls or suppress calls. 

The generated proactive modules have been tested on version 89 of the Xposed framework which requires root privileges to be installed. There are other implementations of Xposed, such as VirtualXposed\footnote{\url{https://virtualxposed.com}}, that do not require root permissions. It is part of our future work to test the integration with these implementations of Xposed.

In addition to tools for developers of proactive modules, we also provide an app that can be used by end users to enable and disable proactive modules for those apps they deem as unreliable. Our current implementation allows direct deployment of the generated proactive modules from Eclipse to Android devices. In a more sophisticated view of this process, proactive modules could be published online, and end-users could use the app to decide what modules to download and install. 
\section{Evaluation}\label{sec:evaluation}
This section summarizes the main results we obtained with Proactive Libraries. Additional details are available in~\cite{Riganelli:TAAS:PL:2019}. 

To evaluate Proactive Libraries, we selected a set of Android apps that have been reported to be potentially affected by API misuses~\cite{Liu:ResourceLeaks:ISSRE:2016} and 3 faulty apps identified in GitHub for a total of 15 real-world Android apps and possible 27 misuses. 

We first identified the usage policies relevant to the APIs involved in the reported misuses and then we defined the corresponding enforcement strategies. To identify the policies, we exploited the information about the use of the Android API reported in other papers~\cite{Liu:ResourceLeaks:ISSRE:2016,Wu:Callback:TSE:2016} and official guidelines~\cite{AndroidAPI_2017}. 

We then executed the test cases that should produce the misuses, and checked whether the misuses have been automatically detected and healed by the proactive modules. In none of the cases the proactive modules fail to heal an execution that violates an API usage policy. This result confirms the effectiveness and suitability of proactive modules to enforce correctness policies. 

The overhead introduced with the proactive modules does not introduce significant differences in the execution time of the apps. This has been confirmed with an ANOVA test that revealed no significant differences (with a significance level equals to $0.05$) between the apps with and without the enforcers running. Considering all 15 apps we have an average overhead of 2\%, even if in several cases (7 out of 15 apps) we have multiple modules active at the same time (from 2 to 6 active modules simultaneously).

We also conducted a human subject study to investigate if the modeling and code generation tool is actually useful compared to implementing Xposed modules from scratch. In particular, we involved three developers who graduated from our department in the implementation of proactive modules. 
When using our development environment, all the subjects produced correct enforcers in less than 1 hour, with the most experienced user completing the task faster than all the other subjects. On the contrary only the most experienced user produced correct enforcers in less than three hours when using Xposed only.

\section{Related Work}\label{sec:related}

Failures caused by API misuses are very popular~\cite{Amani:MUBench:MSR:2016}. These failures have been proven to be pervasive in software systems and different automated approaches are available to detect different types of misuses~\cite{Mariani:BCT:TSE:2011,Wasylkowski:MiningTemporalSpec:ASE:2009,Li:PRMiner:ESECFSE:2005}. However, it is still impossible to completely prevent API misuses, due to limitations of the available techniques that can only detect specific classes of faults.

Even if the concept of proactive library is general, many of the cases studied in the paper are about failures caused by incorrect resource management, such as apps that acquire and release resources according to wrong patterns~\cite{Riganelli:SPE:ResourceLeak:2019,Azim_Towards_2014,Li:GreenProgramming:GREENS:2014,Wu:Callback:TSE:2016,Liu:ResourceLeaks:ISSRE:2016}. Although some of these problems may be discovered with ad-hoc tests and static analysis techniques~\cite{Azim_Towards_2014,Wu:Callback:TSE:2016,Liu:ResourceLeaks:ISSRE:2016}, it is generally difficult to eliminate resource management problems, covering every possible situation, also considering the problem of fragmentation that affects Android~\cite{Wei:AndroidFragmentation:ASE:2016}.

Techniques for avoiding and mitigating the impact of failures at runtime have been studied in many different contexts~\cite{Magalhaes_SSH_2015,Seiger2016,Dai_SHD_2009}. However, only a few limited solutions have been designed to address an environment with limited resources such as a mobile device.

The first results in the Android domain were mainly related to dynamic patch injection, automatic suppression of faulty functionality, and healing of data loss problems. Dynamic patch injection can be used to quickly deploy fixes in apps~\cite{Mulliner_PST_2013, Zhang_AAG_2014}. However, since patches are produced offline, healing is only achieved with the intervention of the app developer, similarly to program repair techniques~\cite{Gazzola:Repair:TSE:2017,Ginelli:APR:2021,Mobilio:FILO:2020}. This mechanism can be useful to solve important vulnerabilities, but it cannot be used for immediate and automatic healing of failing executions. 

Automatic suppression mechanisms can be used to automatically detect crashes and avoid future occurrences of the same crashes by bypassing the execution of features that caused the crash~\cite{Azim_Towards_2014}. This approach may be useful in preventing further crashes, but does not help with fixing problems.

Healing of data loss problems provides a strategy to prevent any data loss due to an incorrect implementation of mechanisms to suspend and recover the execution of the apps~\cite{Riganelli:HealingDataLos:IWSF:2016,Riganelli:MSR:DLBenchmark}. 

Compared to these techniques, proactive libraries provide a design solution that is complementary to mechanisms such as dynamic patch injection, and potentially more general than approaches that address specific classes of failures, such as data loss problems.
\section{Conclusions}\label{sec:conclusion}
We presented the novel concept of Proactive Libraries, which combine a regular reactive library with multiple proactive modules that can monitor the execution and enforce the satisfaction of correctness policies about how to use the API methods at runtime. 

We have implemented a tool that allows developers of proactive modules to design enforcement models according to usage policies and to automatically generate the corresponding proactive modules. These modules are then deployed on the device and the users, using a specifically developed app, decide whether to use them in their own applications.

We used Proactive Libraries to enforce the correct behaviour in 15 apps. The results that we obtained show that proactive libraries can efficiently enforce the specified API usage policies.


\bibliographystyle{ACM-Reference-Format}
\bibliography{main}

\end{document}